\documentclass[aps,pra,twocolumn,showpacs,groupedaddress,letterpaper]{revtex4-1}

\usepackage{graphicx}
\usepackage{mdframed}
\usepackage{framed}
\usepackage{indentfirst}
\usepackage{natbib}
\usepackage{amsmath,amssymb}
\usepackage{subfigure}
\usepackage{enumitem}
\usepackage{epstopdf}
\usepackage{multirow}

\begin{document}

\title{Students' epistemologies about experimental physics: Validating the Colorado Learning Attitudes about Science Survey for Experimental Physics}


\author{Bethany R. Wilcox}
\affiliation{Department of Physics, University of Colorado, 390 UCB, Boulder, CO 80309}

\author{H. J. Lewandowski}
\affiliation{Department of Physics, University of Colorado, 390 UCB, Boulder, CO 80309}
\affiliation{JILA, National Institute of Standards and Technology and University of Colorado, Boulder, CO 80309}

\begin{abstract}
Student learning in instructional physics labs represents a growing area of research that includes investigations of students' beliefs and expectations about the nature of experimental physics.  To directly probe students' epistemologies about experimental physics and support broader lab transformation efforts at the University of Colorado Boulder (CU) and elsewhere, we developed the Colorado Learning Attitudes about Science Survey for Experimental Physics (E-CLASS).  Previous work with this assessment has included establishing the accuracy and clarity of the instrument through student interviews and preliminary testing.  Several years of data collection at multiple institutions has resulted in a growing national data set of student responses.  Here, we report on results of the analysis of these data to investigate the statistical validity and reliability of the E-CLASS as a measure of students' epistemologies for a broad student population.  We find that the E-CLASS demonstrates an acceptable level of both validity and reliability on measures of, item and test discrimination, test-retest reliability, partial-sample reliability, internal consistency, concurrent validity, and convergent validity.  We also examine students' responses using Principal Component Analysis and find that, as expected, the E-CLASS does not exhibit strong factors (a.k.a. categories).  
\end{abstract}

\maketitle

\section{\label{sec:intro}Introduction}

The Physics Education Research (PER) community has a growing body of research dedicated to investigating students' attitudes and epistemologies such as what it means to know, learn, and do physics.  This attention to students' epistemologies stems, in part, from research demonstrating that students' beliefs and expectations about the nature of doing and knowing physics can be linked to both their decision to pursue physics (i.e., retention and persistence) and their content learning in a physics course \cite{lising2005epistemology, hammer1994epist}.  To further investigate these links, researchers have developed several assessments designed to directly measure students' epistemologies and expectations (E\&E) both about physics specifically \cite{adams2006class,redish1998mpex, elby2001ebaps} and the nature of science more generally \cite{halloun1996vass, abd2001vnos, chen2006vose}.  

Until recently, the available physics-specific E\&E surveys have focused on assessing students' epistemologies in the context of instruction in lecture courses.  However, laboratory courses also offer significant and potentially unique opportunities for students to engage in the core practices and ideas of physics.  Indeed, developing students' E\&E has been called out as an important goal of laboratory science courses by multiple national organizations including the American Association of Physics Teachers \cite{AAPT2015guidelines, feynman1998goals}, the National Research Council \cite{nrc2003bio}, and the President's Council of Advisors on Science and Technology \cite{olson2012excel}.  These calls emphasize that effective laboratory instruction should help students develop expert-like habits of mind, experimental strategies, enthusiasm, and confidence in research.  

\begin{figure}
\begin{minipage}{0.95\linewidth}
   \vspace{1mm}\flushleft {\bf Calculating uncertainties usually helps me under-\\stand my results better.} \\
   \hspace{31mm}Strongly  \hspace{23mm} Strongly\\
   \hspace{31mm}disagree \hspace{2mm}1 \hspace{2mm}2 \hspace{2mm}3 \hspace{2mm}4 \hspace{2mm}5 \hspace{3mm}agree \\
   \flushleft
   \begin{minipage}{0.515\linewidth}
      \flushleft\emph{What do YOU think when doing experiments for class?}\vspace{2mm} \\
      \emph{What would experimental physicists say about their research?}
   \end{minipage}
   \begin{minipage}{0.3\linewidth}
      \includegraphics[width=22mm]{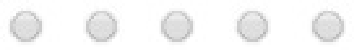}\vspace{5mm}
      \includegraphics[width=22mm]{radio.eps}\vspace{4mm}
   \end{minipage}
\end{minipage}
\caption{An example item from the E-CLASS.  Students are asked to rate their agreement with the statement from their own perspective and that of an experimental physicist.  See Supplemental Material for a list of all item prompts. }\label{fig:eclassExample}
\end{figure}

To support ongoing, nation-wide initiatives to improve laboratory instruction within physics based on these recommendations, researchers at the the University of Colorado Boulder (CU) recently developed the Colorado Learning Attitudes about Science Survey for Experimental Physics (E-CLASS) \cite{zwickl2014eclass}.  E-CLASS is a 30 item, Likert-style survey designed to measure students' epistemologies and expectations about experimental physics (see Supplemental Materials for a list of all item prompts).  Items on the E-CLASS feature a paired question structure in which students are presented with a statement and asked to rate their level of agreement both from their personal perspective and that of a hypothetical experimental physicist (see Fig.\ \ref{fig:eclassExample}) \cite{gray2008class}.  The instrument was developed to target explicit learning goals articulated as part of laboratory course transformations taking place at CU \cite{zwickl2013adlab}.  

The development and initial validation of the E-CLASS included both iterative faculty review and student think-aloud interviews \cite{zwickl2014eclass}.  Twenty-three physics experts reviewed and responded to the E-CLASS survey.  These responses were used to confirm that the instrument effectively targeted the desired learning goals and determine the consensus, expert-like response.  Additionally, 42 interviews were conducted in which students completed the survey while explaining their reasoning out-loud.  The E-CLASS was modified based on these interviews in order to ensure the prompts were clear and correctly interpreted.  For example, the prompt, ``What do YOU think when doing experiments for class?'' was reworded from the original, simpler version, ``What do YOU think?''  The addition of the phrase, ``when doing experiments for class,'' was motivated by evidence from interviews that students with prior research experience would sometimes pull from their personal research experiences, rather than their laboratory course experience, when responding to the survey \cite{zwickl2014eclass}.  

The current version of the E-CLASS has now been administered as an online pre- and post-survey for five semesters at CU and multiple other institutions across the US.  In this article, we build on the initial validation of the E-CLASS through analysis of this national data set in order to establish the full statistical validity and reliability of the instrument for a broad student population.   After a general overview of the data collection and scoring of the E-CLASS (Sec.\ \ref{sec:methods}), we present the results of a detailed statistical analysis of students' responses including, test and item scores (Sec.\ \ref{sec:descriptives}), reliability (Sec.\ \ref{sec:reliability}), validity (Sec.\ \ref{sec:validity}), and principal component analysis (Sec.\ \ref{sec:pca}).  We end with a discussion of limitations and future work (Sec.\ \ref{sec:discussion}).  Throughout this article, we will focus exclusively on analyses related to establishing the statistical validity of the E-CLASS.   Analysis of the data to address broader research questions (e.g., impact of different pedagogical techniques on E-CLASS scores) will be the subject of future publications.

\section{\label{sec:methods}Methods}

In this section, we describe the methods used for collection, scoring, and analysis of student responses to the E-CLASS.  We also present the general demographic breakdown of our final dataset.

\subsection{\label{sec:administration}Data Sources}

All data for this study were collected between January 2013 and June 2015, and, in all courses, the E-CLASS was administered online.  The survey was hosted exclusively at CU, and all student responses were collected directly by the CU research team without needing to be collected by the course instructor first.  To administer the E-CLASS, participating instructors completed a Course Information Survey \cite{eclasswebsite} in which they provided basic information about their course, institution, and pedagogy.  After completing the Course Information Survey, instructors received a link to the online pre- and post-instruction E-CLASS to distribute to their students.  In most cases, the pre- and post-survey remained open for the first and last seven days of the course respectively.  

\begin{table}
\caption{Number of institutions of different types for which we have either pre- \emph{or} post-test responses, or matched pre- \emph{and} post-test responses to the E-CLASS.}\label{tab:instTypes}
   \begin{tabular}{ l r r r r r }
     \hline
     \hline
       & 2-year & 4-year & Master's & Ph.D. \hspace{5mm}& Total\\
       &\hspace{1mm} college &\hspace{1mm} college &\hspace{1mm} granting &\hspace{1mm} granting & \\
     \hline
     All & 0 & 25 & 4 & 20 & 49 \\
     Matched \hspace{1mm}& 0 & 23 & 4 & 17 & 44 \\
     \hline
     \hline
   \end{tabular}
\end{table}

Instructors were recruited to administer the E-CLASS in their courses through a variety of methods including presentations at national professional meetings and emails to professional list-serves.  Information on the instrument was also available on the E-CLASS website \cite{eclasswebsite}, providing an additional avenue for interested instructors to learn about the survey.  Over the five semesters of data collection, we aggregated student responses to the pre- and/or post-instruction E-CLASS from 80 distinct courses spanning 49 institutions including CU.  We have matched pre- and post-data from 71 of these courses spanning 44 institutions.  These institutions include a variety of different types from four-year colleges to Ph.D. granting institutions (see Table \ref{tab:instTypes}).  Additionally, several institutions administered the E-CLASS multiple times in the same course during the 5 semesters of data collection resulting in a total of 103 separate instances of the E-CLASS in our matched set and 114 in the overall pre and post data sets.  Moreover, these courses span the full space of introductory and advanced labs.  Table \ref{tab:courseTypes} shows the breakdown between first-year and beyond-first-year courses in the matched data set.  

\begin{table}[b]
\caption{Number of first-year and beyond-first-year courses in the matched data set.  The number of students in the beyond-first-year courses is smaller in part because of the smaller class sizes typical of more advanced physics labs.  The number of separate instances of the E-CLASS accounts for courses that administered E-CLASS more than once in the 5 semesters of data collection.   }\label{tab:courseTypes}
   \begin{tabular}{ l r r r }
     \hline
     \hline
       & Distinct & Separate & \hspace{1mm} Number of \\
       &\hspace{1mm} courses &\hspace{1mm} instances & Students  \\
     \hline
     First-year & 35 & 49 & 2433  \\
     Beyond-first-year \hspace{1mm}& 36 & 54 & 1158  \\
     \hline
     \hline
   \end{tabular}
\end{table}

\begin{table}
\caption{Number of students in the final pre-, post-, and matched data sets.  Note, gender data was collected only on the post-instruction E-CLASS, and a small number of students opted not to provide their gender. }\label{tab:dems}
   \begin{tabular}{ l r r r}
      \hline
      \hline
       &\hspace{2mm} Total N &\hspace{2mm} Female &\hspace{2mm} Male \\
      \hline
      Pretest & 5658 & - & - \\
      Post-test & 4732 & 35\% & 62\% \\
      Matched Pre \& Post \hspace{4mm}& 3591 & 36\% & 62\% \\
      \hline
      \hline
   \end{tabular}
\end{table}

Certain student responses were eliminated from the final pre- and post-survey data sets because they were identified as invalid.  For a response to be considered valid, the student must must: (1) include at least two of the three student identifiers (first name, last name, and student ID number), (2) respond to multiple survey questions, and (3) respond appropriately to a filtering question.  This filtering question asks students to select the option `Agree' (not `Strongly Agree') and was included in the survey to help eliminate responses from students who randomly selected answers without reading the questions.  Valid student responses were then matched pre to post using student ID number or, when student ID matching failed, first and last name.  

The final breakdown of valid student responses overall and by gender is given in Table \ref{tab:dems}.  Table \ref{tab:dems} does not include a breakdown of the racial demographics of our sample because these data were not collected.  Starting in Fall 2015, the post-instruction E-CLASS will collect data on race of the participants.  Data on students' current major was also collected on the post-survey.  Table \ref{tab:major} reports the breakdown of students by major in the matched data set.  Here, engineering physics majors are included in the `Physics' category and undeclared majors are included in the `Non-science' category.  

\begin{table}[b]
\caption{Breakdown of students by major in the matched data sets (N=3591).  Note, `Physics' includes both physics and engineering physics majors, and `Non-science' includes both declared non-science majors and students who are open option/undeclared.  The exact distribution of majors in any specific course varies significantly based on the type and level of the course. }\label{tab:major}
   \begin{tabular}{ l r r r r}
      \hline
      \hline
       &Physics& \hspace{2mm} Engineering& \hspace{2mm} Other Science& Non-\\
       & & (Non-physics)& \& Math & \hspace{2mm} science \\
      \hline
      Percent & 23\% & 29\% & 40\% & 7\% \\
      \hline
      \hline
   \end{tabular}
\end{table}

 To get an idea of the overall response rate, we compare the number of E-CLASS responses to the number of students enrolled in the course.  Total enrollment was reported by the instructor on the Course Information Survey; however, this number represents only a rough estimate of the actual enrollment because many instructors complete the survey prior to the start of term when the number of students may still fluctuate.  Moreover, the reported enrollment is likely an over estimate as it reflects the initial enrollment of the course and does not take into account those students who drop the course during the term.  The overall, average response rate for our sample was 75\% for the pretest and 64\% for the post-test.  Response rates for individual courses varied significantly.  Factors that may have impacted the response rate include: the level of incentive provided by the instructor (i.e., normal course participation credit, extra credit, or no credit for completion), when the instructor distributed the link, and how the instructor framed the activity to the students.  

In addition to students' responses to the E-CLASS, we also collected student grade data from two semesters of the four core physics laboratory courses at CU in order to address the relationship between E-CLASS scores and laboratory course grade (Sec.\ \ref{sec:convergent}).  These courses span both the lower- and upper-division level, and, in all cases, students were awarded only participation credit for completing the survey.  Final letter grades were collected for all students who agreed to release their grade data.  Only students with matched post-instruction E-CLASS scores and final course grades were included in the grade analysis ($N=873$).

\subsection{\label{sec:scoring}Scoring}

Students' numerical E-CLASS scores are determined only by their responses to the prompt targeting their personal beliefs, rather than their prediction of what an experimental physicist would say (see Fig.\ \ref{fig:eclassExample}).  Moreover, for the purpose of validating the survey, we will focus exclusively on students' pre- and post-instruction scores rather than their shifts from pre to post.  The motivation for this is to minimize the potential for confounding the validation of the survey with the impact of instruction.  This also allows individual implementations of the pre- or post-instruction E-CLASS to provide valid and comparable results that can be used, for example, as baseline data.  

For scoring purposes, students' responses to each 5-point Likert item are condensed into a standardized, 3-point scale in which the responses `(dis)agree' and `strongly (dis)agree' are collapsed into a single category.  Responses are then given a numerical score based on whether they are consistent with the consensus expert response.  The expert response can be either agree or disagree depending on particular item \cite{zwickl2014eclass}, and thus, student responses to individual items are coded simply as favorable (+1), neutral (0), or unfavorable (-1).  The collapsing of the 5-point scale to 3-points is common in analysis of Likert-style items and is motivated, in part, by the inherently ordinal, rather than interval, nature of the Likert response scale \cite{lovelace2013attitudes}.  The use of the 3-point scale is also supported by previous literature suggesting that the threshold between `Agree' and `Strongly agree' is not always consistent between individual students or groups with different cultural backgrounds \cite{halloun2001vass}.  

Previous literature on E\&E surveys, has often defined a students' overall score as the fraction of items to which they responded favorably \cite{adams2006class}.  This 2-point scoring scheme treats neutral and unfavorable responses the same.  However, we consider the distinction between a neutral and unfavorable response to be valuable and argue that the overall score should include this distinction.  Thus, students' overall E-CLASS score is given by the sum of their scores on the individual items on the 3-point scale described above.  This results in a range of possible scores from -30 to 30 points.  To explore the impact of different scoring conventions, we performed all of the analyses described in Sec.\ \ref{sec:results} using both the 2-point and 3-point scoring schemes and found that the two schemes resulted in the same conclusions with respect to the validity and reliability of the E-CLASS.

\subsection{\label{sec:analysis}Analysis}

There are multiple potential approaches to the analysis of multiple-choice tests \cite{ding2009mcAnalysis}.  We utilized two of these approaches here: Classical Test Theory (CTT) \cite{engelhardt2009ctt} and Principal Component Analysis \cite{oRourke2013factor}.  CTT establishes the validity and reliability of a multiple-choice assessment based on the assumption that a students' score is composed solely of both their true score along with some random error \cite{ding2009mcAnalysis}.  Validation of an assessment via CTT involves analysis of student responses to calculate multiple test statistics and evaluation of these statistics relative to accepted thresholds (see Sec.\ \ref{sec:results}).  One major drawback to CTT stems from the fact that all test statistics are calculated using student responses and, thus, are dependent on the specific population of students.  As a consequence, there is no guarantee that test statistics calculated for one student population (e.g., undergraduate students) will hold for another population (e.g., high school students).  For this reason, scores on assessments validated through the use of CTT can only be clearly interpreted to the extent that the student population matches the population with which the assessment was validated \cite{wallace2010irt}.  

One alternative to CTT for establishing the statistical validity of a multiple-choice assessment is Item Response Theory (IRT).  IRT addresses many of the shortcomings of CTT by providing a method for producing population independent estimates of both item and student parameters \cite{wallace2010irt, ding2009mcAnalysis}.  In the simplest IRT models, a student's performance on an item is assumed to depend only on their latent ability and the item's difficulty.  For test items that fit this model, all item and student parameters calculated via IRT are independent of both population and test form \cite{ding2009mcAnalysis, baker2001irt}.  However, IRT models also assume that the assessment is unidimensional (i.e., designed to measure a single construct).  The E-CLASS, on the other hand, was explicitly designed to target multiple, potentially non-overlapping aspects of students' beliefs about the nature of experimental physics including: modeling, statistical analysis, troubleshooting, etc \cite{zwickl2014eclass}.  For this reason, we have chosen to utilize CTT, rather than IRT, to establish the statistical validity of the E-CLASS.  

Literature on existing E\&E surveys often groups questions into categories (a.k.a.\ factors) and reports students' scores in each of these categories.  In some cases, this categorization was based on \emph{a priori} criteria imposed by the developer (e.g., Ref.\ \cite{redish1998mpex}).  In other cases, the categorization was based on statistical analyses such as factor analysis, which identified statistically robust categories of questions (e.g., Refs.\ \cite{adams2006class, lindstrom2012fa}).  However, items on the E-CLASS were not specifically developed to match a specific categorization scheme, but rather to target a wide range of individual learning goals.  In other words, the E-CLASS was not designed with specific categories in mind; thus, we have no \emph{a priori} cause to believe that the E-CLASS would exhibit strong factors.  

In order to determine whether or not the E-CLASS can be adequately characterized by a relatively small number of question groups, we utilize Principal Component Analysis (PCA).  PCA is a data reduction technique used to reduce the number of variables in a data set while still capturing a significant fraction of the variance \cite{lindstrom2012fa, oRourke2013factor}.  PCA is typically used in data sets where there is reason to believe that there is significant redundancy among the variables \cite{oRourke2013factor} and uses an inter-item correlation matrix to identify groups of items that appear to vary together.  We performed a PCA on students' responses to the both the pre- and post-instruction E-CLASS from the matched data set using the statistical software package R \cite{r2015website}.  

\section{\label{sec:results}Results: Statistical Validity}

This section presents evidence for the statistical validity and reliability \cite{engelhardt2009ctt} of the E-CLASS using the pre, post, and matched data sets described in Sec.\ \ref{sec:administration}.  The results of the principal component analysis will be discussed in Sec.\ \ref{sec:pca}.

\subsection{\label{sec:descriptives}Test \& Item Scores}

\begin{figure}
   \includegraphics[width=0.9\linewidth]{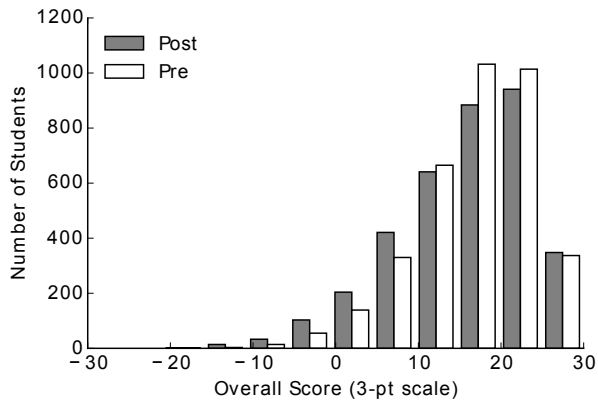}
\caption{Distribution of students pre- and post-instruction E-CLASS scores for students with matched pre- and post-scores ($N=3591$).  The average E-CLASS score was $16.5 \pm 0.1$ points ($\sigma = 6.8$) for the pretest, and $15.4 \pm 0.1$ points ($\sigma = 7.9$) for the post-test.  The difference between the two distributions is statistically significant (Mann-Whitney U \cite{mann1947mwu}, $p<< 0.05$).}\label{fig:overallDist}
\end{figure}

As described in Sec.\ \ref{sec:scoring}, a students' overall E-CLASS score is given by the sum of their scores on each of the 30 items where each item is scored on a 3-point scale (favorable = +1, neutral = 0, unfavorable = -1).  Fig.\ \ref{fig:overallDist} shows the distribution of overall E-CLASS scores for the matched data set ($N=3591$).  The difference between the pre- and post-distributions is statistically significant (Mann-Whitney U \cite{mann1947mwu}, $p<< 0.05$), though the effect size is small (Cohen's $d = 0.14$ \cite{cohen2013d}).  The students' overall performance on the E-CLASS can also be summarized by looking at the average fraction favorable relative to the average fraction unfavorable.  Table \ref{tab:overallFrac} reports these statistics for the matched pre- and post-instruction E-CLASS.  

\begin{figure*}
\includegraphics[width=0.95\linewidth]{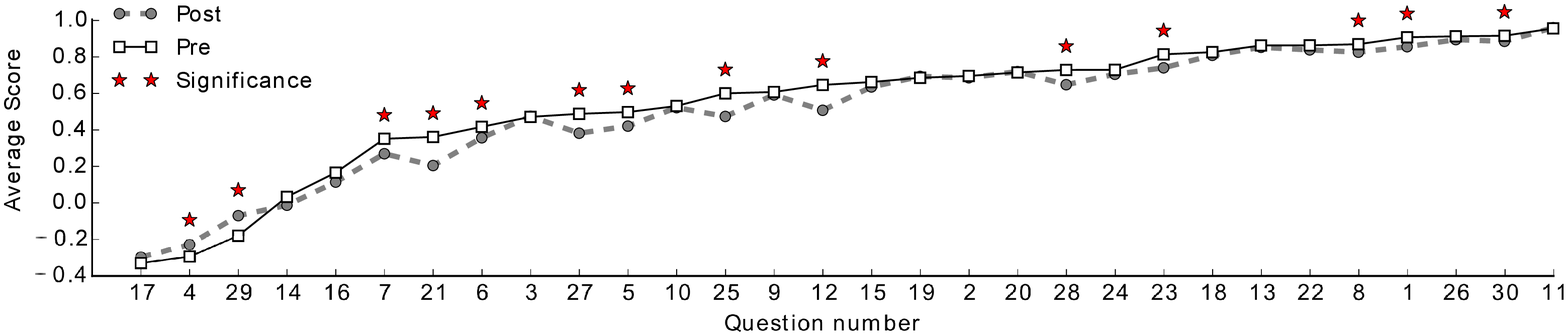}
\caption{Average item scores using the 3-point scoring scale for each of the 30 E-CLASS items.  Items are sorted in ascending order by scores on the pre-instruction E-CLASS.  Statistically significant differences between the pre- and post-instruction averages are indicated by a red star (Holm-Bonferroni corrected $p<0.05$).  See Supplemental Material for a list of individual prompts. }\label{fig:Idiff}
\end{figure*}

In addition to looking at the overall E-CLASS score, we can also examine students' scores on each item individually.  A sorted graph of the average pre- and post-test scores for each of the 30 E-CLASS items is given in Fig.\ \ref{fig:Idiff}.  Fig.\ \ref{fig:Idiff} also highlights questions for which the difference between the pre- and post-instruction scores is statistically significant.  Statistical significance was determined at the $\alpha = 0.05$ level, and p-values were corrected for multiple-testing effects using the Holm-Bonferroni method \cite{holm1979hb}.  While there is no standard criteria for acceptable item scores on Likert items \cite{ding2006bema}, it is typically argued that ideal item scores should be targeted towards maximizing the potential discriminatory power of each item and the test as a whole \cite{doran1980measurement}.  Fig.\ \ref{fig:Idiff} shows that the average item scores for the 30 E-CLASS items all fall between -0.4 and 0.96.  This wide range suggests that the E-CLASS is capturing a significant amount of variation in students epistemologies and expectations about experimental physics.  

\begin{table}[b]
\caption{Average fraction of items with favorable and unfavorable responses for the matched pre- and post-instruction E-CLASS. } \label{tab:overallFrac}
\begin{tabular}{l l r r r}
    \hline
    \hline
    & & \hspace{1mm} Average &  \hspace{1mm} Standard & \hspace{1mm} (Standard \\
    & & & Error & Deviation) \\
    \hline

   Favorable & Pre & 0.69 & 0.002 & (0.15) \\
    & Post \hspace{1mm} & 0.68 & 0.003 & (0.16) \\
   \hline
   Unfavorable \hspace{1mm} & Pre & 0.15 & 0.002 & (0.10) \\
    & Post & 0.17 & 0.002 & (0.12) \\
   \hline
   \hline
\end{tabular}
\end{table}

\subsection{\label{sec:reliability}Reliability}

The reliability of an assessment relates to the instrument's ability to measure the targeted construct(s) \emph{consistently}.  In this section, we present measures of several different aspects of the reliability of the E-CLASS including: test-retest reliability, time-to-completion reliability, partial-sample reliability, internal consistency, and testing environment (i.e., in-class versus online) reliability.

\subsubsection{\label{sec:trR}Test-retest reliability}

The test-retest reliability of an assessment looks at the stability of students' scores.  In other words, if students were to take the test twice with no intervention, they should, in theory, get the same score.  Straightforward measures of test-retest reliability are difficult to achieve for a number of logistical reasons including maturation and retest effects.  One proxy that can be used to establish the test-retest reliability of the E-CLASS is the stability of pre-instructions scores from semester to semester of the same course.  As the students in a particular course are drawn from the same population each semester, we can reasonably expect that the pre-instruction scores for that course should remain relatively stable over time.  

During the 5 semesters of data collection, we collected matched pre-post data for two or more semesters of 16 different courses at 11 institutions.  Of these, only 5 courses showed statistically significant differences between the average pre-instruction scores between semesters.  In all cases, these differences stemmed from a single semester with anomalously high or low scores, and the effect sizes ranged from small (two courses, $d=0.2$) to large (three courses, $d>0.5$).  However, while we expect the pre-instruction population of a course to be relatively stable, it is also reasonable to expected that there would be small, legitimate variations in this population from cohort to cohort.  The small number of statistically significant variations detected in our pretest data is consistent with this, thus supporting the test-retest reliability of the E-CLASS.  

Another proxy for the test-retest reliability of the E-CLASS looks at whether students' scores shift from the beginning to the end of the semester when they are not enrolled in a laboratory course.  To investigate the stability of students scores under the null condition, we administered the E-CLASS pre- and post-surveys to one instance of the middle-division classical mechanics course at CU ($N=49$).  Given the standard course sequence at CU, the majority of the students in this course are physics and engineering physics majors and are not taking a physics laboratory course at the same time.  To identify any off sequence students, the post-instruction E-CLASS for this course was modified to ask students to report which (if any) laboratory courses they were taking that semester.  

We collected 38 matched responses to the E-CLASS from the classical mechanics course, 10 of whom were concurrently enrolled in a physics laboratory course.  The remaining students ($N=28$) had an average pre-instruction score of $17.6\pm1.6$ ($\sigma=8.5$) and a post-instruction score of $17.6 \pm1.5$ ($\sigma=7.8$).  The difference between these scores is not statistically significant, suggesting that these students' epistemologies did not shift over the course of a semester in which they were not enrolled in a laboratory course.  Alternatively, the ten students who were concurrently enrolled in a physics laboratory course showed a slight decrease in their scores ($22.2$ points to $21.6$ points) over the semester.  With only ten students, this decrease is not statistically significant; however, the direction and magnitude of the shift are consistent with that in the overall population (Fig. \ref{fig:overallDist}).  This finding also preliminarily supports the test-retest reliability of the E-CLASS; however, additional data from other courses and institutions will be necessary to robustly establish the stability of students' E-CLASS scores under the null condition.  

\subsubsection{\label{sec:ttcR}Time-to-completion Reliability}

It is also worth examining whether there is any correlation between the amount of time each student spends on the survey and their final score.  In most cases, the E-CLASS was administered as an out-of-class online assignment, and start and stop times for each student were automatically collected by our online system for all semesters except one.  The time elapsed is defined as the amount of time between when the student first clicked on the survey link to their final submission.  Between these two times, however, a student may, for example, step away from the computer or close out the survey and start again at a later time.  For this reason, the time elapsed does not necessarily represent the amount of time the student actually spent completing the survey.  The median time elapsed was 8 minutes for the pretest and 12 minutes for the post-test.  The correlation between overall score and time elapsed was $|r| < 0.02$ for both pre- and post-tests.  This correlation is neither statistically nor practically significant, suggesting that there was no link between time-to-completion and E-CLASS score for the matched data set.  

\subsubsection{\label{sec:psR}Partial Sample Reliability}

For any assessment with less than 100\% response rate, it is also important to keep in mind the potential for selection effects when interpreting students scores.  Because students essentially self select into those who respond and those who do not, the sample of respondents may end up over-representing certain subpopulations (e.g., high performers) and under-representing others (e.g., low performers).  To examine one aspect of the partial-sample reliability of the E-CLASS, we compare the average overall score for the matched data set with the average overall score of students who took only either the pre- or post-survey.  We found no statistically significant difference in average pre-instruction scores for the matched and unmatched data sets.  This same difference for the post-instruction scores was statistically significant ($p << 0.05$) with the unmatched students scoring slightly below the matched, but the effect size was small ($d = 0.2$).  

In addition to looking at the partial-sample reliability of the matched data relative to the unmatched data, we can also examine the population of respondents and non-respondents relative to other measures of student performance.  In order to establish the convergent validity of the E-CLASS (see Sec.\ \ref{sec:convergent}), we collected course grade data for all students in CU's four, core laboratory courses over two semesters during which we also administered the E-CLASS.  From these data, we can compare the average course grade of students who completed both the pre- and post-instruction E-CLASS ($N=875$) with those who completed only one or neither ($N=459$).  We found that students in the matched data set had an average laboratory course grade of $3.3$ (on a 4pt scale) while the average for those who competed only one or neither was $2.7$.  The difference in final grade between the matched and unmatched students is statistically significant and represents a relatively large effect size ($d=0.7$).  This suggests that low response rates likely result in an under-representation of lower performing students.  

This selection effect is unsurprising, but is an important factor for instructors and researchers to keep in mind when interpreting the results of the E-CLASS for courses with lower response rates.  The over-sampling of the higher performing students may suggest that results from courses with low response rates should be interpreted as a best case scenario snapshot of the overall student population's epistemologies.

\subsubsection{\label{sec:consistency}Internal Consistency}

Another aspect of test reliability is the consistency of students' scores on arbitrary subsets of items.  For a unidimensional test, Cronbach's alpha is a measure of this type of internal consistency \cite{cortina1993ca}.  Cronbach's alpha can be loosely interpreted as the average correlation between all possible split-half exams.  For the purposes of low-stakes testing of individuals, a minimum value of $\alpha = 0.8$ is considered acceptable \cite{engelhardt2009ctt}.  For the matched data set, we found $\alpha = 0.76$ for the pre-survey and $\alpha = 0.83$ for the post-survey.  However, the interpretation of Cronbach's alpha generally assumes a unidimensional test targeting a single construct and multidimensionality within the assessment will tend to drive alpha downward \cite{cortina1993ca}.  The E-CLASS, alternatively, was not designed to be unidimensional, but rather to target multiple, potentially overlapping, aspects of students' ideas; thus, Cronbach's alpha represents a conservative estimate of the internal consistency of the E-CLASS.  The potential multidimensionality of the E-CLASS will be discussed in more detail in Sec.\ \ref{sec:pca}.

\subsubsection{\label{sec:teR}Testing Environment Reliability}

In the majority of courses, students complete the E-CLASS outside of class time; however, a subset of instructors have students complete the assessment during class time, usually in order to increase participation.  Giving the E-CLASS in-class is most viable in courses that take place in a computer lab in which there is one student per computer.  To investigate students' scores in different testing environments (i.e., in-class vs. out-of-class), we administered the E-CLASS during class time to the first-year lab course at CU in the fall of 2015 ($N=521$).  We compared the average overall score from this course with that from the same course in the 5 previous semesters ($N=1568$).  We found a small (Cohen's $d=0.16$) but statistically significant (Mann-Whitney U, $p=0.003$) increase in students' pre-instruction E-CLASS scores from the full pre-instruction data set for this course.  However, the difference between pre-instruction averages was not statistically significant for the subset of the pre-instruction data set for this course that had matched post-instruction responses.  This suggests that for this population of students taking the E-CLASS during class time had, at most, a small positive impact on their performance, though that increase did not persist in the matched sample.  While this finding preliminarily supports the testing environment reliability of the E-CLASS, additional data from other institutions and courses will be necessary to clearly establish the impact of testing environment on students' scores.  

\subsection{\label{sec:validity}Validity}

\begin{figure*}
\includegraphics[width=0.95\linewidth]{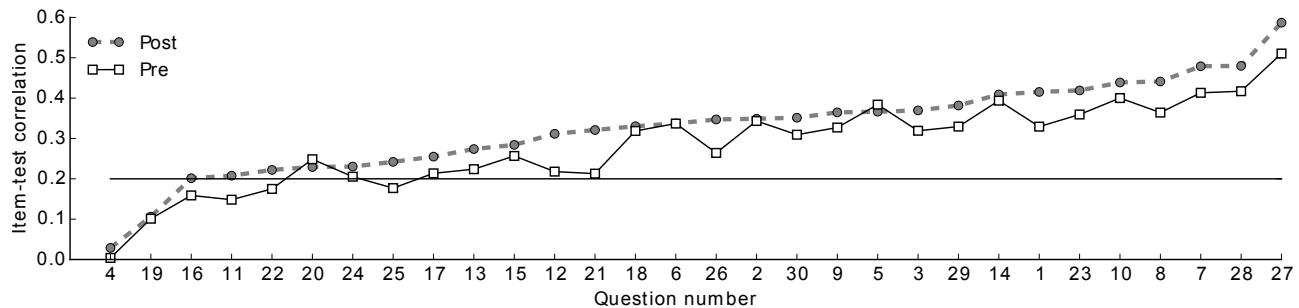}
\caption{Item-test correlations using the 3-point scoring scale for each of the 30 E-CLASS items.  Items are sorted in ascending order by post-instruction $r$-values.  The standard threshold for an acceptable item-test correlation is noted as a horizontal line at $r=0.2$.  See Supplemental Material for a list of individual prompts. }\label{fig:Idisc}
\end{figure*}

The validity of an assessment relates to the instrument's ability to \emph{accurately} measure the targeted construct(s).  In this section, we present measures of several different aspects of the validity of the E-CLASS including: discrimination (whole-test and by-item), concurrent validity, and convergent validity.  

\subsubsection{\label{sec:discrim}Discrimination}

To examine the ability of the E-CLASS overall to discriminate between students, we used Ferguson's delta \cite{ding2009mcAnalysis}.  Roughly speaking, Ferguson's delta looks at how well scores are distributed over the full range of possible point values, in this case, -30 to 30 points.  Delta can range from [0,1] and anything above 0.9 indicates good discriminatory power \cite{ding2006bema}.  For both the pre- and post-instruction E-CLASS, we found $\delta > 0.98$, well above the standard threshold.  

We also examined the discrimination of each individual item by looking at student's scores on individual items relative to their performance on the E-CLASS as a whole.  Fig.\ \ref{fig:Idisc} shows item-test correlations for each of the 30 E-CLASS items.  Here, we adopt the standard threshold for an item-test correlation for dichotomously scored items, $r > 0.2$ \cite{ding2006bema}.  The majority of the E-CLASS items and the average post-instruction item-test correlation ($r=0.33$) fall above this threshold.  These results support the conclusion that the E-CLASS demonstrates adequate whole-test and item discrimination.  

\subsubsection{\label{sec:concurrent}Concurrent Validity}

\begin{table}[b]
\caption{Average overall scores for FY and BFY courses. Differences between FY and BFY averages are statistically significant (Mann-Whitney U, $p<<0.05$).  } \label{tab:concurrentV1}
\begin{tabular}{l l r r r r}
    \hline
    \hline
    & & Average &  \hspace{1mm} Standard & \hspace{1mm} Significance & \hspace{1mm} Effect \\
    & & (points) & Error & & Size \\
    \hline
   FY & Pre & 15.8 & 0.1 & \multirow{2}{*}{$p<<0.05$} & \multirow{2}{*}{0.2} \\
   ($N=2433$) & Post & 14.4 & 0.2  \\
   \hline
   BFY \hspace{1mm} & Pre & 17.7 & 0.2 & \multirow{2}{*}{$p=0.3$} & \multirow{2}{*}{0.04} \\
   ($N=1158$) & Post & 17.5 & 0.2  \\
   \hline
   \hline
\end{tabular}
\end{table}

Concurrent validity examines the extent to which E-CLASS scores are consistent with certain expected results.  For example, it is reasonable to expect that students' scores will vary between different levels of courses.  In particular, first-year courses are often service courses catering primarily to engineering, biology, or non-science majors, rather than physics majors.  Thus, the learning goals of these courses may be less aligned with some of the learning goals targeted by E-CLASS, which were developed in collaboration with physics faculty to capture their desired learning outcomes for physics students in their upper-division lab courses \cite{zwickl2014eclass}.  Moreover, we anticipated that students in the higher level courses would have more expert-like responses, either due to selection effects or the cumulative impact of instruction, or some combination of both.  To investigate this, we divided the students into two subgroups composed of those in first-year (FY) labs ($N=2433$) and those in second-, third-, and fourth-year labs ($N=1158$), which we will refer to as the beyond-first-year (BFY) labs.  This division between FY and BFY labs is consistent with the classification of courses used in the community of laboratory instructors and provides a relatively clear distinction between courses that is applicable both at CU and other institutions.  

\begin{table}[b]
\caption{Average overall scores for first-year physics and non-physics majors. Differences between the majors are all statistically significant (Mann-Whitney U, $p<<0.05$).  } \label{tab:concurrentV2}
\begin{tabular}{l l r r r}
    \hline
    \hline
    & & \hspace{1mm} Average &  \hspace{1mm} Standard & \hspace{1mm} Effect \\
    & & (points) & Error & Size \\
    \hline
   Physics & Pre & 19.4 & 0.4 & \multirow{2}{*}{0.1} \\
   ($N=215$) & Post & 18.6 & 0.5  \\
   \hline
   Non-Physics \hspace{1mm} & Pre & 15.5 & 0.1 & \multirow{2}{*}{0.2} \\
   ($N=2218$) & Post & 14.0 & 0.2 \\
   \hline
   \hline
\end{tabular}
\end{table}

The average pre- and post-instruction E-CLASS scores for the FY and BFY courses is given in Table \ref{tab:concurrentV1}.  The difference between the E-CLASS scores of the FY and BFY courses is a statistically significant ($p<<0.05$) for both the pre- and post-surveys.  Thus, students in the BFY courses both begin and end the semester with more expert-like views than students in the FY course.  In addition to expecting students in the BFY courses to score higher on the E-CLASS, we might also anticipate that physics majors will score higher than non-physics majors even within the FY courses.  Here, we include engineering physics majors in our population of physics majors.  Average pre- and post-instruction E-CLASS scores for FY physics and non-physics majors are given in Table \ref{tab:concurrentV2} and show that FY physics majors both begin and end with significantly more expert-like views than non-majors.  Both of these findings are consistent with expectations and support the concurrent validity of the E-CLASS.

\subsubsection{\label{sec:convergent}Convergent Validity}

Convergent validity looks at whether scores on an assessment correlate with other, related student outcomes \cite{adams2011development}.  For conceptual assessments, convergent validity is typically established relative to students' final course grades.  However, for E\&E assessments like the E-CLASS, it is reasonable to ask if we expect the same level of correlation with course performance \cite{redish1998mpex,hammer1989two}, particularly given that promoting expert-like attitudes and beliefs is rarely an \emph{explicit} goal of physics courses.  Of the available E\&E assessments, only the VASS (Views About Science Survey) has published results reporting a modest but significant correlation ($r\sim0.3$) between VASS score and final course grades in high school and college physics courses \cite{halloun2001vass}.  For the purposes of convergent validity, we will focus exclusively on students' post-instruction E-CLASS scores as these are the most likely to be consistent with final course grades.  Grade data was collected from two semesters of the four core laboratory courses at CU, and students were assigned a standard grade point value for each letter grade (i.e., $A=4.0$, $A-=3.7$, $B+=3.3$, $B=3.0$, etc.).  Aggregating across all students in all courses (N=873), we found an overall correlation coefficient of $r=0.04$ between final course grade and post-instruction E-CLASS score.  This correlation is neither practically or statistically significant ($p=0.2$).  

However, consistent with the previous section, which showed higher overall scores in BFY courses, it is also reasonable to expect that this correlation might vary between courses.  To investigate this potential variability across courses, we divided the students into two subgroups composed of those in the FY lab (N=717) and those in the second-, third-, and fourth-year (BFY) labs ($N=156$).  For students in the FY lab, the correlation between overall E-CLASS score and final grade is small and not statistically significant ($r=0.004$, $p=0.9$).  However, for the BFY labs, this correlation increased to $r=0.17$ and is statistically significant ($p=0.03$).  This correlation, while still weak, is similar in magnitude to the correlations reported between CLASS/MPEX scores and conceptual learning gains as measured by conceptual assessments such as the Force Concept Inventory \cite{perkins2005class, pollock2005reforms, kortemeyer2007mpex}.  We are not arguing that the relationship between E-CLASS scores and final grades is a causal one; however, these results do suggest that the link between course performance and epistemological stance is stronger in more advanced lab courses than in the first-year lab.  

Overall, these findings suggest that, even for BFY courses, E-CLASS scores are not good predictors of students' course grade (or vice versa).  One interpretation of this is that students' epistemological knowledge and beliefs are not being effectively captured by their final course score.  This finding is also consistent with results from the CLASS that have demonstrated neutral or even negative shifts in CLASS scores from courses with significant conceptual learning gains \cite{pollock2006change}.

\section{\label{sec:pca}Principal Component Analysis}

In addition to looking at students' scores overall and by item, existing E\&E surveys often examine students' aggregate scores on smaller groups of items.  These item groupings are typically either based on an \emph{a priori} categorization created by the survey developer \cite{redish1998mpex}, or empirically derived using a statistical data reduction technique \cite{adams2006class}.  As previously discussed, the E-CLASS was not developed to match any particular \emph{a priori} categorization of questions; however, it is still possible that different questions on the E-CLASS target the same latent variable.  In order to determine if this is the case, we utilize Principal Component Analysis (PCA).  

PCA is a type of exploratory factor analysis that attempts to reduce the number of variables in a data set by examining inter-item correlations in order to identify groups of items that appear to vary together \cite{lindstrom2012fa, oRourke2013factor}.  We first performed a PCA on students' responses to the post-instruction E-CLASS from the matched data set.  The initial exploratory PCA produced 30 components (a.k.a. factors) along with associated eigenvalues.  To determine how many of these components to extract, we adopted the Guttman-Kaiser criterion \cite{yeomans1982guttman}, which states that all components with eigenvalues greater than 1  should be kept.  This criteria resulted in 7 components that together explained 45\% of the overall variance in the survey.  However, it is generally accepted that to sufficiently represent the overall data set, the retained components should account for at least 70\% of the variance.  Meeting this threshold for the E-CLASS data would require retaining 16 of the 30 components.  This factor of 2 decrease does not represent a useful reduction in the number of variables in the data set.  

To determine if the factors identified in the post-instruction data were robust, we performed the same PCA on student responses to the pre-instruction E-CLASS.  We found that the there was little overlap between the items that made up specific factors in the pre-survey data compared to those from the post-survey.  This result, along with the fact that the results of the PCA accounted for less than half the overall variance, suggests that the E-CLASS does not exhibit a clear factorization.  This lack of strong factors is not surprising given that PCA is used to identify cases in which there are multiple items targeting a single latent variable.  The E-CLASS, on the other hand, was designed to target a relatively large number of distinct, though potentially overlapping, learning goals.  The PCA suggests that, consistent with this design, the E-CLASS does not appear to contain groups of items targeting a single latent variable.  Given this result, we strongly recommend that instructors do not only focus on their students overall E-CLASS score as it does not represent students' performance around a single well-defined construct.  Rather, instructors should examine their students responses to the questions individually with a particular focus on those questions that are most aligned with their learning goals for that course.

\vspace{12pt}
\section{\label{sec:discussion}Summary \& Future Work}

We previously created an attitudinal survey -- known as the E-CLASS -- targeting students' epistemologies and expectations about the nature of experimental physics.  Prior work established the content validity of this assessment through expert review and student interviews.  To build on this initial validation, we collected student responses to the E-CLASS from roughly 80 courses spanning approximately 45 institutions.  Analysis of these data supports the statistical validity and reliability of the E-CLASS as measured by multiple test statistics including, item and whole-test discrimination, internal consistency, partial-sample reliability, test-retest reliability, and concurrent validity.  A principal component analysis of these data also showed that the E-CLASS does not exhibit robust factors that can be used to accurately and reliably describe students' responses using a smaller number of statistically consistent groups of questions.  

Future work will include analysis of our growing national data set of student responses to the E-CLASS to answer broader research questions regarding students' ideas about the nature of experimental physics.  For example, this data set includes some longitudinal data that could begin to provide insight into how students' epistemologies change over the course of their undergraduate career.  Additionally, the course information survey, completed by all instructors prior to using the E-CLASS, collects information on both pedagogy and learning goals (e.g., learning physics content vs. developing lab skills).  These data can be used to determine if certain pedagogical strategies or learning goals promote more expert-like epistemologies and expectations.  Future research will also include investigating if and how instructors use their students' E-CLASS results to inform their instruction or course transformation efforts.

\begin{acknowledgments}
This work was funded by the NSF-IUSE Grant DUE-1432204.  Particular thanks to Benjamin Zwickl for his work on the initial development and validation of the E-CLASS.  Additional thanks to the members of PER@C for all their help and feedback.  
\end{acknowledgments}

\bibliography{master-refs-ECLASS-6-15-v2}

\end{document}